# Tree based classification of tabla strokes

Subodh Deolekar[1,*], Siby Abraham[2]

[1]Department of Computer Science, University of Mumbai, Mumbai 98, India. subodhdeolekar@gmail.com
[2]Department of Maths & Stats, G. N. Khalsa College, University of Mumbai, India. sibyam@gmail.com

**Corresponding author and address:**
**Subodh Deolekar**
Department of Computer Science, University of Mumbai, Mumbai 98, India.
Email id: subodhdeolekar@gmail.com

**ABSTRACT**

The paper attempts to validate the effectiveness of tree classifiers to classify tabla strokes especially the ones which are overlapping in nature. It uses decision tree, ID3 and random forest as classifiers. A custom made data sets of 650 samples of 13 different tabla strokes were used for experimental purpose. 31 different features with their mean and variances were extracted for classification. Three data sets consisting of 21361, 18802 and 19543 instances respectively were used for the purpose. Validation has been done using measures like ROC curve and accuracy. The experimental results showed that all the classifiers showing excellent results with random forest outperforming the other two. The effectiveness of random forest in classifying strokes which are overlapping in nature is done by comparing the known results of that with multi-layer perceptron.

## I. INTRODUCTION

Classification is the process which assigns a specific item to one of the categories or classes specified based on its features or properties. In machine learning, classification is considered to be a task to predict the value of one or more outcomes. The real task in classification is to find a relationship between features and its associated classes. There are various categories of classification which include linear classifiers, support vector machines, quadratic classifiers, kernel estimation, decision trees and neural networks[1]. Among these, tree based classifiers are commonly used for developing prediction algorithms for a target variable[2]. Such classifiers construct a root node with a population of branches which comprise of internal nodes and leaf nodes.

Tree classifiers aim to partition dataset into groups of similar nature. They are said to be very effective methods of supervised learning, which lead to generate unique solutions. In cases where impurity exists in the data and where there are traces of one class oversteps into another, tree classifiers are best suited. Unlike linear models, they map non-linear relationships quite well. Following are some of the advantages of using tree classifiers:

- *Comprehensive behavior:* Tree classifiers are the best predictive models as they extensively examine each possible outcome. The partition of data is done in a much deeper way as compared with other classification techniques.

- *No need for tuning of parameter set:* Normalization or scaling of the parameter set can be avoided. Where most of the classification models fail to handle non-linearity of parameters, tree classifiers outperform with such data.
- *Easy to interpret:* Tree classifiers make a clear distinction with all possible solutions which are represented by different nodes. A graphical view of classification based on rules and parameter set makes decision making less ambiguous.
- *Easily deal with outliers:* Tree classifiers are flexible in handling data items with some missing feature values. Also, splitting of sub-trees is based on split range and not based on absolute values which depict non-sensitivity towards outliers.

Tree algorithms have been extensively used for classification of various tasks in diverse domains. It include characterizing smoking patterns of older adults[3], analyzing students' achievements in distance-learning mode to improve online teaching[4], identifying core factors of production control in agricultural investment decisions[5], understanding behavioral patterns of different kinds of astronomical objects[6], analyzing financial data[7], text classification[8] and many more. In the domain of music, there are various applications of tree classifiers which consider classification of various instruments[9] and speech/music classification and segmentation[10].

This paper is an attempt to measure the effectiveness of tree algorithms to classify strokes of musical instrument called Tabla. The Indian percussion instrument, which is used for solo performance as well as an accompanied, often reports of cases where impurities exist in its data. In addition, one can find genuinely perceived instances where one stroke seemed to sidestep or overstep into territory of another stroke making classification difficult. This makes classification of tabla strokes a fit case for a tree classifier. Though literature reports of a similar work of classification of tabla strokes using other classifiers[11], the proposed work is different in the sense that it faces the inherent constraint of classification of tabla strokes head on using tree classifiers.

## II. TABLA INSTRUMENT

Tabla plays a very important role in Hindustani music which is most popular and cultural oriented tradition of Indian music. It is used to provide rhythmic pattern in music compositions along with other Indian percussion instruments like Pakhwaj, Mridangam, Dholak and Dholki. Because of refined tonal quality and sophistication, tabla came in forefront as a solo performing instrument leaving behind all other instruments.

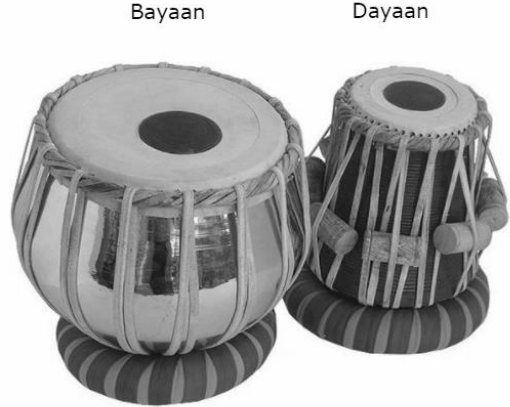

**Fig. 1.** Tabla set

Tabla set, which is a combination of two drums namely *bayaan* (left hand) and *dayaan* (right hand), as shown in figure 1, produces strokes called *bol*s. *Bol*s, which are mnemonic syllables, can be produced using different techniques based on diverse musical traditions[12]. Table 1 gives the list of basic strokes played on *dayaan* and *bayaan* individually and collectively:

**Table 1.** List of Basic Tabla Strokes.

| *DaayaanBols* | *BaayaanBols* | *Both Together* |
|---|---|---|
| Ge (गे / घे) | Na (ना) | Dha (धा) |
| Ka (क / की) | Tu (तू) | Dhin (धीं) |
|  | Ti (ति) | Tin (तीं) |
|  | Ta (ट) |  |
|  | N (न) |  |
|  | T (त / र) |  |
|  | Tra (त्र) |  |
|  | Din (दिं) |  |

Transcription of tabla strokes is said to be a challenging task for computer scientists[13]. Unlike western rhythmic instruments (such as drums or congo), where characteristics of strokes are easily detected, tabla strokes are hard to detect. Tabla produces long and resonant *bols* which will overlap with successive *bols*. Also, few bols which are used to represent rhythmic strokes depend on the specific context and may vary accordingly. There exist *bols,* which are different but sound similar. For example, *Ti* and *Ta* are two *bols,* which are played with subtle changes in the fingering style but are hardly identified correctly by even experienced listeners.

### III. DECISION TREE ALGORITHMS

Decision tree algorithms are flexible, powerful and offer high performance for prediction problems. Without any pre-defined assumption and controlled parameters, these algorithms are capable of fitting large amount of data. In principle, decision tree algorithms categorize data by considering their attribute values. Each parent node in a tree represents a test on an attribute value and child node represents corresponding classes.

## *Decision Tree*

Decision tree is a supervised learning algorithm which has pre-defined target variables. It works for both categorical and continuous input/output variables. The split of sample at root node or at the internal subsequent nodes is based on characteristics of child node. These characteristics are defined by different variables like entropy and gini index.

In information theory, degree of disorganization in a system is called as entropy. It can also be considered as a degree of randomness of elements or a measure of impurity. Mathematically, it can be calculated as:

$$H(x) = -\sum p(x) \log p(x)$$

where $p(x)$ is probability of target variable x is actually transmitted.

On the other hand, gini index performs only binary splits with categorical targets like 'success' or 'failure'. Higher value of gini represents higher homogeneity. It is defined as:

$$Gini = \sum_{i=1}^{C}(P_i)^2$$

Where C is the number of classes and $P_i$ is the probability of $i^{th}$ target variable for i ∈ {1, 2, …, C}.

## *ID3*

Iterative Dichotomiser 3 (ID3) algorithm classifies data using attribute values. Tree consists of decision nodes and decision leafs which produce a homogeneous result. It is based on top-down greedy search to test each attribute at every node of a tree. It calculates the entropy value for each of the attribute[14]. It, then splits the set into subsets using that attribute for which the information gain is maximum. Information Gain is the entropy of the parent node minus the entropy of the child nodes and is given by:

$$Information\ gain = H(X) - \sum_{i=1}^{C}(P_i * H(X_i))$$

Where $H(X)$ is the entropy of the parent node, $H(X_i)$ is the entropy of the child node and $P_i$ is the probability of $i^{th}$ target variable for i ∈ {1, 2, …, C}, C being the number of classes. In this way, ID3 tree at every stage selects the node that gives the best information gain, the one with least impurity[15].

## *Random Forest*

Random Forest is a classifier which comprises of a set of weak, weakly correlated and non-biased classifiers, namely the decision trees. It has been shown that random forest performs equally well or better than other methods on a diverse set of problems. It has been widely used in classification problems as diverse as bioinformatics[16], medicine[17], transportation safety[18] and customer behavior[19].

Random forest offers a useful feature that improves our understanding of a classification problem under scrutiny. It gives an estimate of the importance of each attribute for final prediction. It is often used for analysis when both classifier and identification of important variables are goals of the study.

Random forest collects votes from different decision trees which are randomly selected from training set data and decides the final class of test data. This is helpful for finding accurate results because a single tree might lead to a noise, but a set of decision trees will reduce the noise.

## IV. METHODOLOGY

The paper proposes a methodology which classifies tabla strokes, wherein exist an overlap between different classes, making classification difficult. The methodology, the overview of which is shown in figure 2, consists of four steps namely pre-processing, feature extraction, classification and authentication.

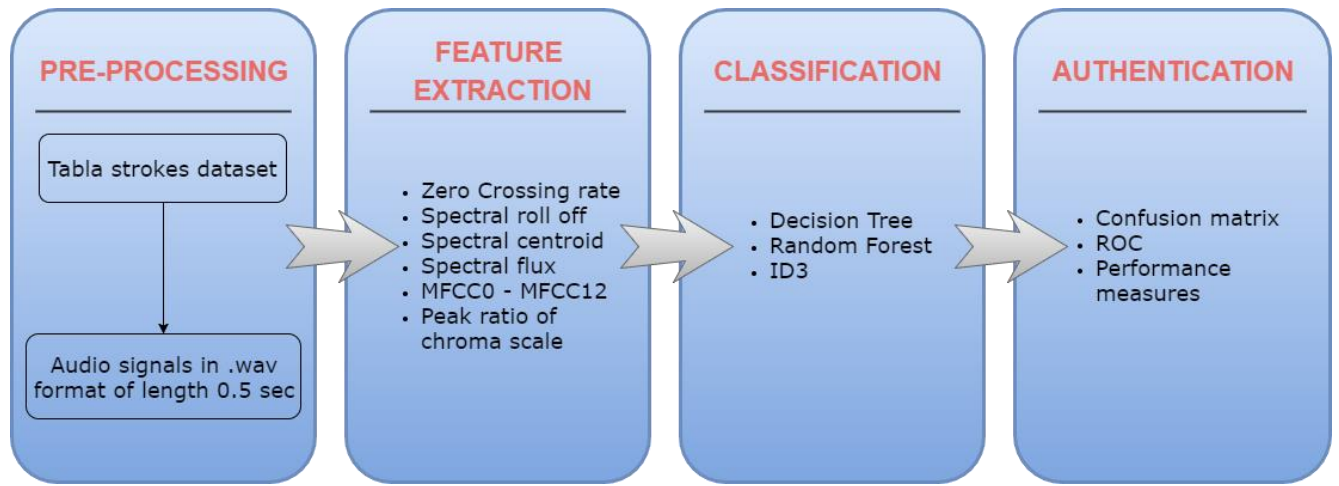

**Fig. 2.** Overview of Proposed methodology

**Pre-Processing**

As most of the publicly available datasets lack authenticity, the methodology uses data sets generated exclusively for this work. Samples of tabla strokes were recorded in the required audio format. These were then clipped to specific time duration to have the same length for each stroke. Initially the duration was fixed to be 0.5 seconds. Those bols, whose resonance do not last for such a long duration were clipped to 0.2 to 0.3 seconds.

A sample of different bols used in the work is shown in waveform in figure 3.

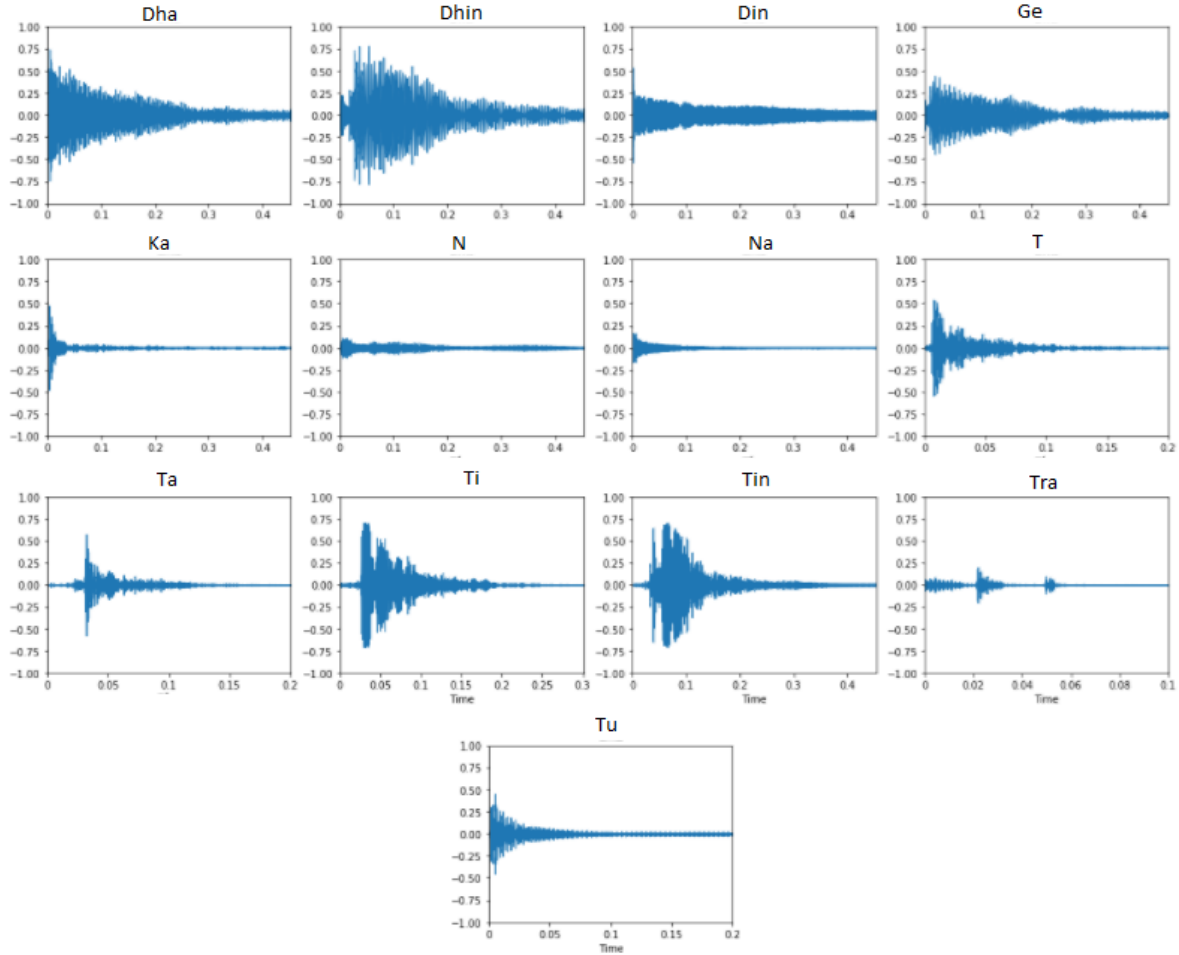

**Fig. 3.** Sample tabla strokes

**Feature Extraction.**

Different sets of spectral and temporal features were considered for analyzing different content of the tabla strokes for the classification purpose[20]. Features like Zero-crossing, Spectral Centroid, Spectral Roll-off, Spectral Flux, Mel-Frequency Cepstral Coefficients (MFCC0-MFCC12), and Chroma frequencies with their mean and standard deviation were extracted from the audio file of individual strokes. A short description of these features is given below:

*Zero Crossing Rate:* It represents the number of times the waveform crosses zero axis. It usually has higher values for highly percussion sounds.

*Spectral Centroid:* It is the weighted mean of the frequencies present in the music piece. The value of it changes according to the accumulation of frequencies.

*Spectral Roll-off:* It is a measure of the shape of a signal. It represents the frequency at which high frequencies decline to zero.

*Spectral flux:* It is a measure of change in the power spectrum of a signal. It is used to determine the timbre of an audio data.

***Mel-Frequency Cepstral Coefficient (MFCC):*** MFCC represents a set of short term power spectrum characteristics of the music piece and has been used in the state-of-the-art recognition and music categorization techniques. Altogether, 13 coefficients from MFCC0 to MFCC12 are identified for this feature.

***Chroma Frequencies:*** Chroma frequency vector discretizes the spectrum into chromatic keys and represents the presence of each key. It provides a robust way to describe a similarity measure between music pieces.

These low level musical features are helpful in the classification process.

**Classification**

We have used three classifiers namely decision tree, ID3 and random forest in order to classify tabla strokes. The choice of tree classifiers is deliberate since it works well with missing attribute values and non-linear working set[21]. For decision tree, the measures used for selecting input variables are entropy and gini and split on each node is binary. ID3 classifier makes use of information gain based on the training samples and builds the tree. For random forest the output is determined based on the majority votes of the trees. As a result random forest uses a large number of trees collectively.

**Authentication**

The methodology has used measures like ROC curve, tree structure based on gini/entropy and accuracy to authenticate the performance of classifiers. Graphical view of the decision tree makes it easy to interpret results along with the other performance measures.

## V. EXPERIMENTAL RESULTS

Detailed experiments were conducted using different tabla strokes. The tabla set used for recording was tuned to C# scale. The original data was comprised of 650 strokes which were split into 50 samples each of 13 tabla strokes recorded from different expert tabla players in order to have diversity. The recording was done using a microphone input in an environment with less noise. After feature extraction, the three datasets of moderate size consisted of 21361, 18802 and 19543 instances respectively were used for classification. For each of the classifier, the data were split into 70% for training and 30% testing. The sample rate was kept as 44100 Hz.

Pre-processing work was done with the help of Audacity[22] software. The features were extracted using MARSYAS (Music Analysis, Retrieval and Synthesis for Audio Signals) framework[23]. The feature values extracted were stored in .csv file for further processing. Python was used to implement tree algorithms. Intel(R) Core i5-CPU with 1.70 GHz with 4 GB RAM was used for the experimental purpose.

*Feature extraction*

The experiments results validated the overlapping nature of original tabla stokes. Features extracted demonstrate this clearly. For example, spectral centroid and zero crossing rate of audio samples of two bols *ti* and *ta* were considered for a representative nature. Figure 4 (a) and (b) shows the power spectrum of these two sample bols and Figure 5 shows the overlap between these two features.

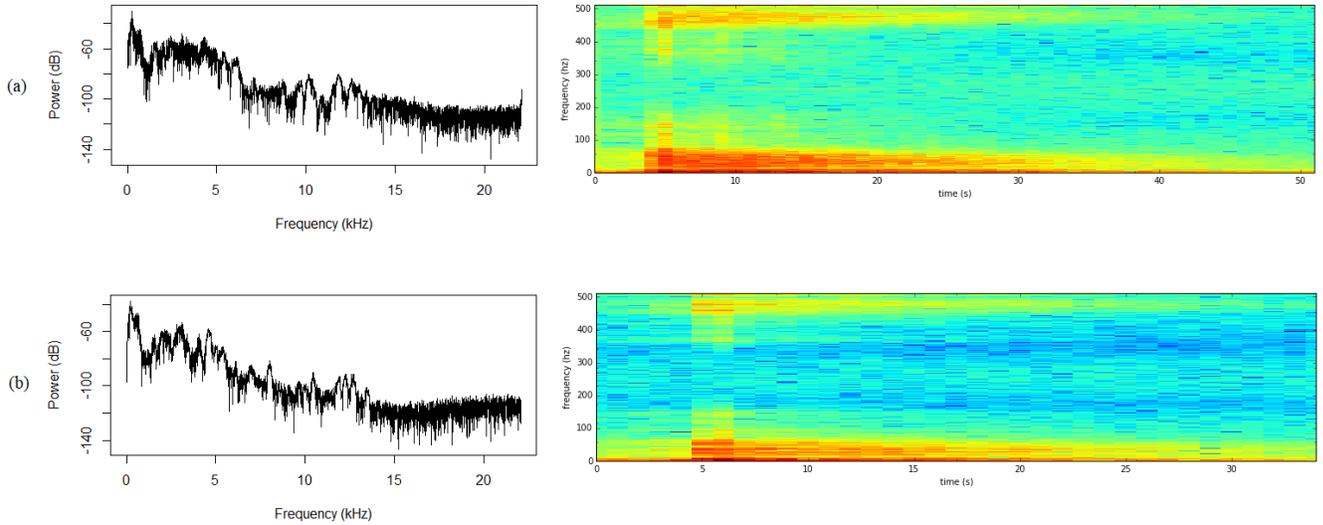

**Fig. 4.** Waveform and Spectrogram of two bols (a) *ti* and (b) *ta* respectively

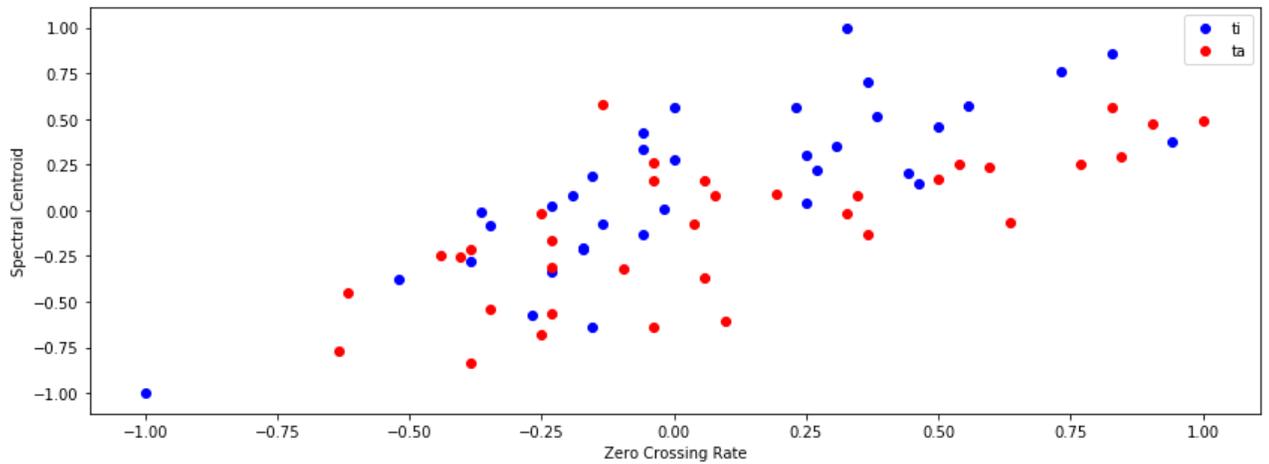

**Fig. 5.** Spectral centroid and zero crossing rate for two bols *ti* and *ta* respectively

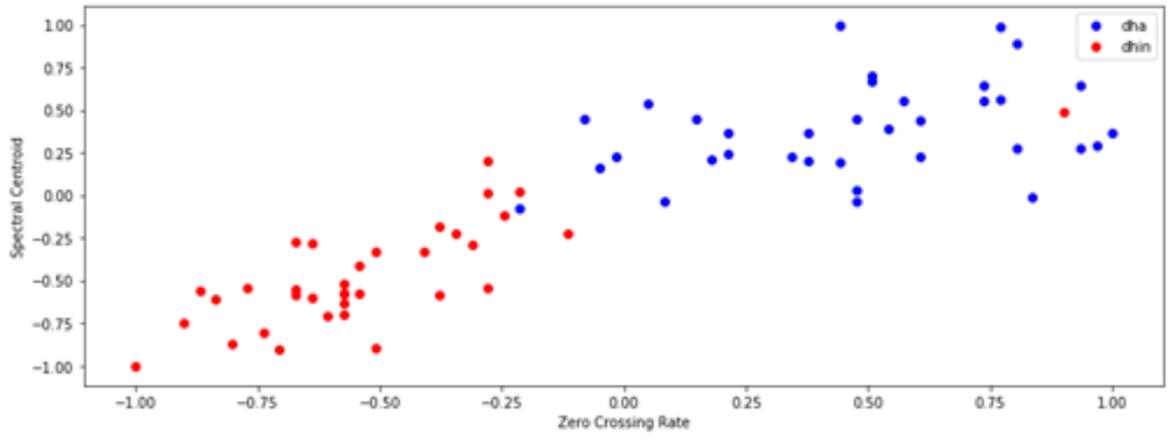

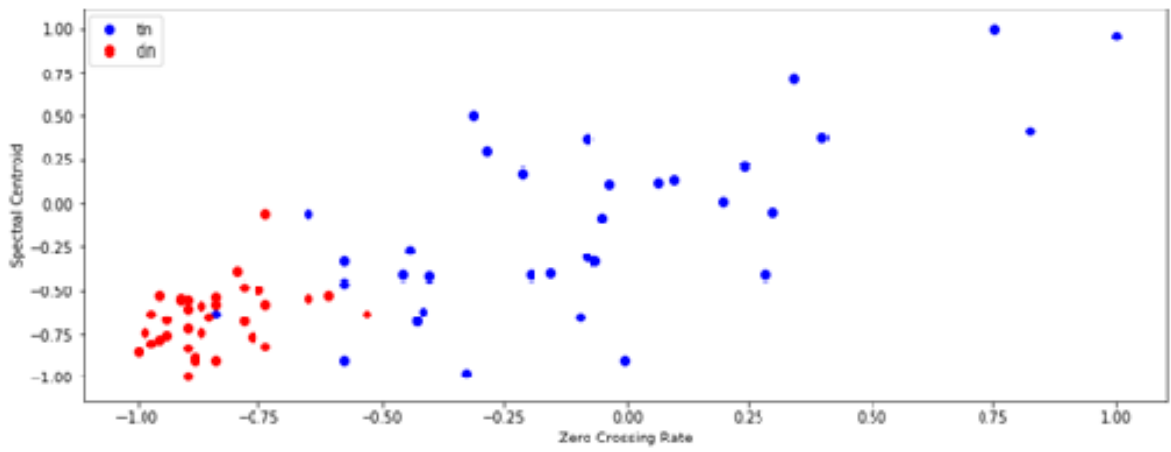

**Figure 6.** Spectral centroid and zero crossing rate for (a) *dha* and *dhin* (b) *tin* and *din* strokes

Similar features were extracted for others bols like *dha, dhin, tin, din* where we found less overlap. Figure 6(a) and 6 (b) shows these less overlapping classes.

*Decision tree representation*

The work described used tree classifiers viz. decision tree, ID3 and random forest on the extracted features set. The Figure 7 shows a sample decision tree generated using gini index criteria and Figure 8 shows the decision tree generated from entropy criteria using the dataset 1in both cases.

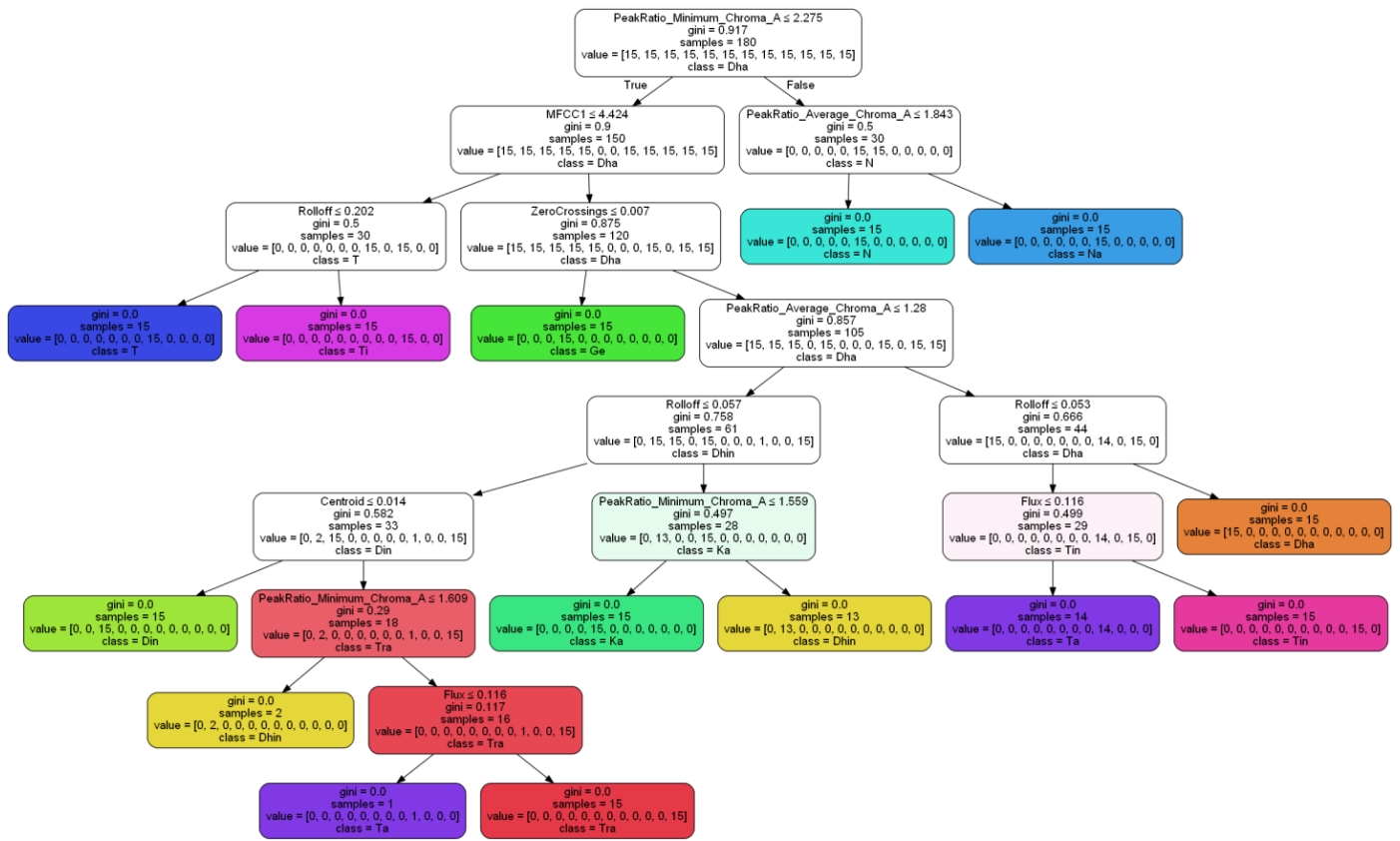

**Figure 7.** Decision Tree predicting the tabla strokes based on gini index value

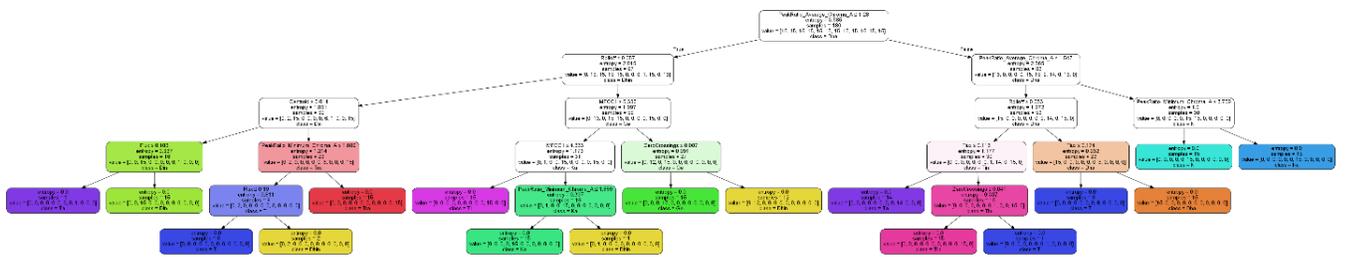

**Figure 8.** Decision Tree predicting the tabla strokes based on entropy value

It can be observed that the decision tree splits the node 2-way. It grows depth wise with gini index and expands depth wise with entropy. That means, Gini index favors the larger partition whereas entropy favors smaller partitions with many distinct values. The value for gini index and entropy becomes zero when all the observations belong to same class label.

In Figure 7 we see the PeakRatio_Minimum_Chroma_A feature used for the initial split with gini index value as 0.917. When we used entropy for the base calculation, it was observed that a wider range of results, whereas the gini index capped at one. In Figure 8 PeakRatio_Average_Chroma_A is chosen for the split and has entropy value as 3.585.

Classifier performance varies depending upon the characteristics of the data used in the classification process. We tried to perform various empirical evaluation strategies to compare classifiers performance.

*ROC*

Receiver Operating Characteristic (ROC) curves has been used to evaluate the tradeoff between true and false positive rates of decision tree classifier. It is considered to be a plot of sensitivity versus 1-specificity for all possible thresholds of different classes.

Figure 9 shows the ROC curve for the Random forest classifier. It is observed that random forest performs well with classifying all the possible tabla strokes with accuracy close to 100%. It was found out that with overlapping bols *ti* and *ta,* which are represented by class 8 and 9, the accuracy was more than 90%.

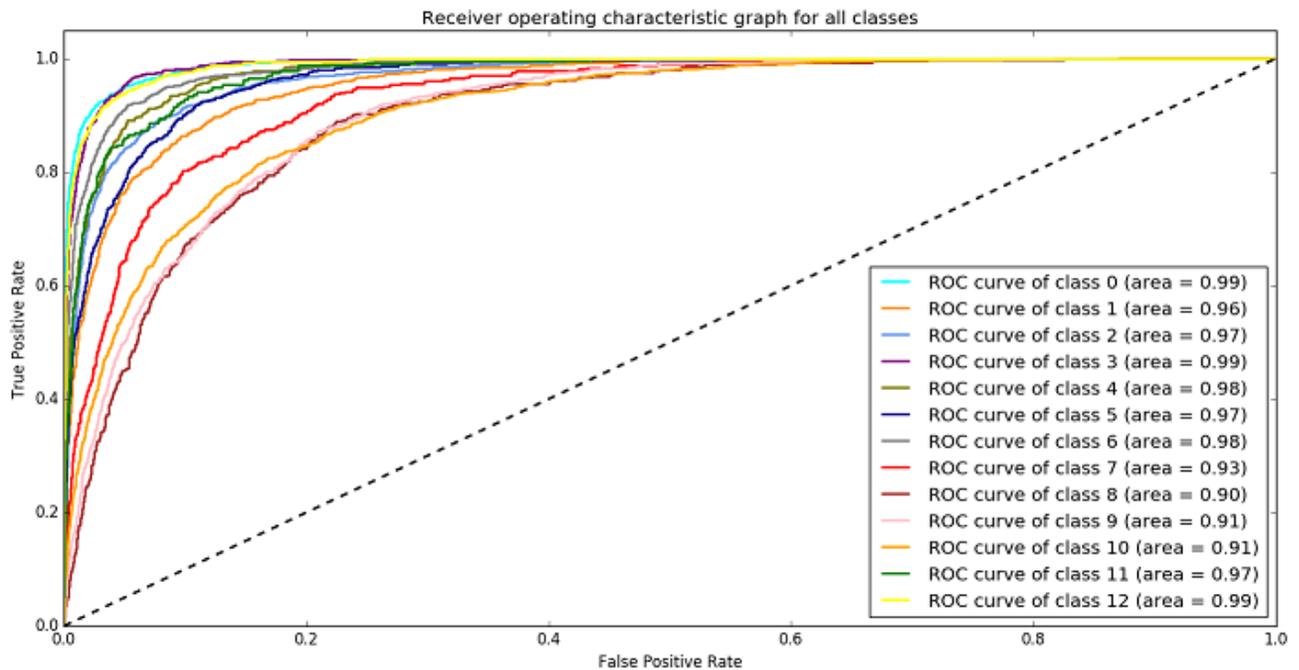

**Figure 9.** ROC Curve for all the classes

*Accuracy measurement*

Experimental results showed that Random Forest outperformed other classifiers in terms of accuracy with all three datasets. While random forest gave an accuracy up to 100%, ID3 (97%) and Decision tree (99%) also have performed well. Figure 10 illustrates these findings.

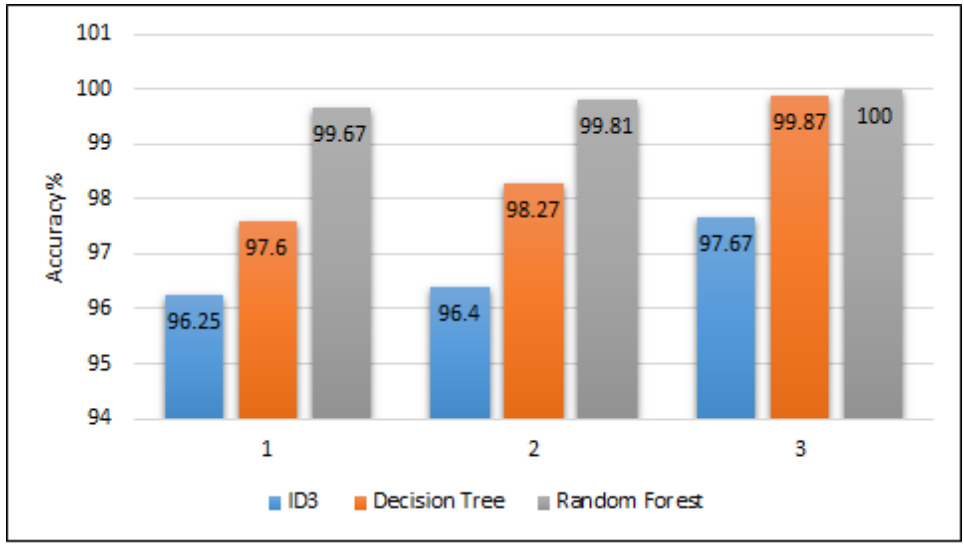

**Figure 10.** Accuracy comparison between ID3, Decision tree and Random forest algorithm

*Accuracy comparison of overlapping strokes*

We tried to compare the results of random forest with earlier work with Multilayer Perceptron (MLP) classifier [11] by considering overlapping tabla strokes of *ti* and *ta*. It was observed that the random forest exhibited a much better accuracy of around 91% while that of MLP was hovering around 80 -82%, showing the effectiveness of tree algorithms, especially the random forest in properly classifying overlapping class instances. Figure 11 gives an idea about the supremacy of random forest as a tree classifier for tabla strokes classification in comparison with the available work in the literature.

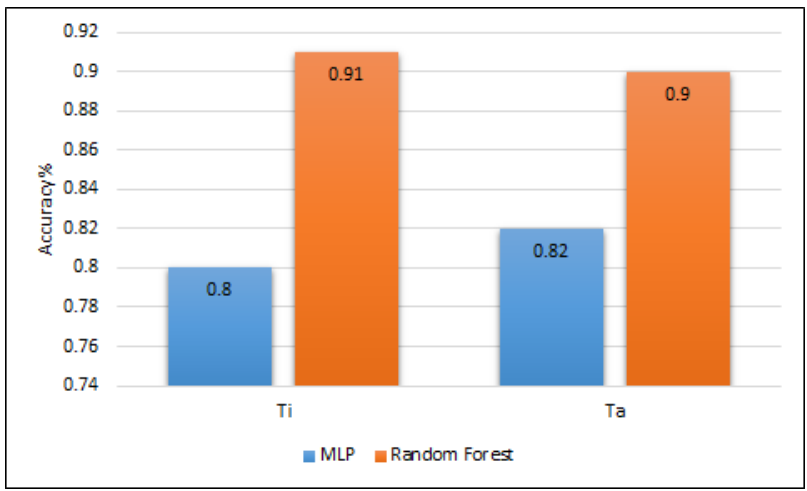

**Figure 11.** Comparison of accuracy of *ti* and *ta* stokes for MLP and Random forest algorithm

All these results shows that Random forest algorithm is very well suited to classification of tabla strokes and works significantly better than other two tree classifiers namely Decision tree and ID3. The performance of random forest in classifying overlapping classes also showed to be better than the one offered by other classifiers.

## Conclusion

The proposed work highlights effectiveness of tree classifiers in classifying tabla strokes. It uses decision tree, ID3 and random forest as classifiers. Detailed experiments were conducted using three different data sets. 31 different features of the strokes with their mean and variances were extracted. The classification results show that random forest outperforms the other two classifiers. The performance of random forest was compared with other known classifier and results showed that random forest is better suited to classify overlapping classes. The work can be extended classify strokes of other percussion instruments also.

## Acknowledgement

We would like to thank Mr Arun Kundekar and his team for their help in recording tabla strokes and offering their expertise in the domain.

## References

1. Sagar S. Nikam, A Comparative Study of Classification Techniques in Data Mining Algorithms, Oriental Journal of Computer Science & Technology, April 2015,Vol. 8.

2. Yan-yan, S. and Ying, L. Decision tree methods: applications for classification and prediction, Shanghai Archives of Psychiatry, 2015, vol. 27, no. 2, pp. 130-135.

3. Sung Seek Moon, Suk-Young Kang, Weerawat Jitpitaklert, Seoung Bum Kim, Decision tree models for characterizing smoking patterns of older adults, Expert Systems with Applications, Volume 39, Issue 1, January 2012, Pages 445-451.

4. Wilton W.T. Fok, Haohua Chen, Jiaqu Yi, Sizhe Li, H.H. Au Yeung, Wang Ying, Liu Fang, Data Mining Application of Decision Trees for Student Profiling at the Open University of China, *2014 IEEE 13th International Conference on Trust, Security and Privacy in Computing and Communications*, Beijing, pp. 732-738.

5. Lu J., Liu Y., Li X., The Decision Tree Application in Agricultural Development. In: Deng H., Miao D., Lei J., Wang F.L. (eds) Artificial Intelligence and Computational Intelligence. AICI 2011. Lecture Notes in Computer Science, vol 7002. Springer, Berlin, Heidelberg.

6. A. Franco-Arcega, L. G. Flores-Flores and R. F. Gabbasov, Application of Decision Trees for Classifying Astronomical Objects, *2013 12th Mexican International Conference on Artificial Intelligence*, Mexico City, pp. 181-186.

7. G. Yu and G. Wenjuan, Decision Tree Method in Financial Analysis of Listed Logistics Companies, *2010 International Conference on Intelligent Computation Technology and Automation*, Changsha, pp. 1101-1106.

8. Srinivasan Ramaswamy, Multiclass Text classification A Decision Tree based SVM Approach, CS294 Practical Machine Learning Project, Citeseer, 2006.

9. Kursa M., Rudnicki W., Wieczorkowska A., Kubera E., Kubik-Komar A., Musical Instruments in Random Forest. In: Rauch J., Raś Z.W., Berka P., Elomaa T. (eds) Foundations of Intelligent Systems. ISMIS 2009. Lecture Notes in Computer Science, vol 5722. Springer, Berlin, Heidelberg.

10. Y. Lavner and D. Ruinskiy, A Decision-Tree-Based Algorithm for Speech/Music Classification and Segmentation, EURASIP Journal on Audio, Speech, and Music Processing, 2009.

11. Deolekar S., Abraham S., Classification of Tabla Strokes Using Neural Network. In: Behera H., Mohapatra D. (eds) Computational Intelligence in Data Mining—Volume 1. Advances in Intelligent Systems and Computing, 2016, vol 410. Springer, New Delhi.

12. David Courtney, Learning the Tabla, Volume 2, M. Bay Publications, 2001, ISBN 0786607815.

13. Chordia, P., Segmentation and recognition of tabla strokes. International Conference on Music Information Retrieval, 2005, 107–114.

14. A. Navada, A. N. Ansari, S. Patil and B. A. Sonkamble, Overview of use of decision tree algorithms in machine learning, *2011 IEEE Control and System Graduate Research Colloquium*, Shah Alam, pp. 37-42.

15. Jin C. De-lin L. Fen-Xiang M. An improved ID3 decision tree algorithm, in the proceeding of 4th International Conference on Computer Science & Education, 2009, 127-130.

16. Qi Y., Random forest for bioinformatics, In: Zhang C., Ma Y. (eds) Ensemble Machine Learning. Springer, Boston, MA, 2012, pp. 307–323.

17. Dubrava S, Mardekian J Sadosky A*, et al*, Using random forest models to identify correlates of a diabetic peripheral neuropathy diagnosis from electronic health record data. Pain Med 2017; 34: 107–15.

18. Saha, D., Alluri, P., and Gan, A., A random forests approach to prioritize Highway Safety Manual (HSM) variables for data collection. J. Adv. Transp., 2016, 50: 522–540.

19. Armando Vieira, Predicting online user behaviour using deep learning algorithms. Computing Research Repository - arXiv.org, 2015, http://arxiv.org/abs/1511.06247.

20. W. Brent, Physical and perceptual aspects of percussive timbre, UCSanDiego Electronic Theses and Dissertations, 2010.

21. Chattamvelli R., Data Mining Methods, Alpha Science International, Oxford, UK. 2009.

22. Audacity® software is copyright © 1999-2017 Audacity Team. Web site: https://audacityteam.org/. It is free software distributed under the terms of the GNU General Public License. The name Audacity® is a registered trademark of Dominic Mazzoni.

23. George Tzanetakis and Perry Cook. 1999. MARSYAS: a framework for audio analysis. *Org. Sound*4, 3, 169-175.